\documentclass[runningheads]{svmult}

\usepackage{makeidx}   
\usepackage{graphicx}  
\usepackage{subeqnar}  
\usepackage{multicol}  
\usepackage{physprbb}  
\usepackage{float}

\newcommand{\beq}{\begin{equation}}
\newcommand{\eeq}{\end{equation}}
\newcommand{\ket}[1]{| {#1} \rangle}
\newcommand{\bra}[1]{\langle {#1} |}
\newcommand{\ave}[1]{\langle {#1} \rangle}
\newcommand{\tave}[1]{\langle\!\langle{#1}\rangle\!\rangle}

\begin{document}
\title*{Kaon Condensation in Neutron Stars}
\author{Angels Ramos\inst{1}
\and J\"urgen Schaffner-Bielich\inst{2}
\and Jochen~Wambach\inst{3}}
\authorrunning{A. Ramos et al.}
\institute{
Departament d'Estructura i Constituents de la Mat\`eria,
Universitat de Barcelona,
Diagonal 647, 08028 Barcelona, Spain
\and
RIKEN BNL Research Center, Brookhaven National Laboratory,
Upton, New York 11973-5000, USA
\and
Institut f\"ur Kernphysik, TU Darmstadt,
Schlossgartenstr. 9, 64289 Darmstadt, Germany}

\maketitle              

\begin{abstract}
We discuss the kaon-nucleon interaction and its consequences for the change of 
the properties of the kaon in the medium. The onset of kaon condensation in neutron stars
under various scenarios as well its effects for neutron star properties are reviewed. 
\end{abstract}


\section{Introduction -- hadrons in dense matter}
\label{sec:intro}

Due to its non-abelian structure, Quantum Chromodynamics (QCD) becomes
very strongly interacting and highly nonlinear at large space-time distances. 
As a consequence, quarks
and gluons condense in the physical vacuum with a gain in condensation
energy density of $\Delta\epsilon_0\sim 500$ MeV/fm$^3$. The condensation of quarks is
associated with the spontaneous breaking of chiral symmetry, an additional
symmetry of the QCD Lagrangian in the absence of (current) quark masses.
This limit is well justified in the up-down quark sector and to a somewhat
lesser extent for the strange quark. There is good evidence that the 
mechanism for spontaneous chiral symmetry breaking is provided by
classical gluon field configurations in euclidean space called 'instantons'.
These provide effective quark-(anti)quark 
interactions which are strong enough to cause a BCS like transition to 
a condensed state of quarks and antiquarks. It
has been shown, that this picture provides an excellent description of
hadronic states and correlation functions \cite{SchaSh} and it is fair
to say that the low-lying hadron spectrum is dominated by spontaneous
chiral symmetry breaking with confinement playing a much lesser role. 
These observations form the basis for discussing the properties of
hadrons, or more precisely hadronic correlation functions, under extreme
conditions in temperature and/or density as encountered in the early
universe or in the interior of neutron stars. It is well established from
lattice QCD that the vacuum undergoes a phase transition
(or at least a sharp cross over) when heated. Chiral symmetry is restored
in this phase accompanied by a nearly vanishing chiral quark condensate. Though
not calculable at present from first principles, the same is expected to 
happen at finite density. Since light hadrons are dynamically driven by
chiral symmetry breaking it is obvious to ask how hadronic properties are
related to the vacuum structure and its changes with temperature and 
density. This is far from trivial and under intense experimental and
theoretical scrutiny at present.

As detailed below the most economical way for treating hadrons in
matter under extreme conditions is to resort to 'effective field theories'
in which hadrons rather than quarks and gluons appear as the fundamental
degrees of freedom. Formally one identifies the pertinent quark currents
$J_{\Gamma_j}(x)=\bar q(x)\Gamma_j q(x), \Gamma_j=1, \gamma_5,\gamma_\mu,..$
with elementary hadronic fields $\phi_i(x)$ for which the most general
effective Lagrangian, consistent with the underlying symmetries and
anomaly structure of QCD, is written down. 
We recall
that the spontaneous breaking of chiral symmetry has two important
consequences. One is the appearance of (nearly) massless Goldstone 
bosons (pions, kaons, eta) and the other the absence of parity doublets 
in the hadron spectrum ($m_\pi\neq m_{f_0}, m_\rho\neq m_{a_1}$ etc).
For the present discussion the first is the most relevant. 
Chiral symmetry does more than just predict the existence of Goldstone
bosons. It also prescribes and severely restricts their mutual
interactions as well as those with other hadrons.
The most rigorous treatment is
in terms of 'chiral perturbation theory' \cite{GaLe} but other
'chiral effective Lagrangians' including e.g.\ vector mesons can be
devised \cite{NK,NK97}.

To elucidate the connection between the vacuum structure (the chiral
condensate) and the properties of light hadrons let us consider a
medium of hadronic matter in thermal and chemical equilibrium. This
is of course well suited for neutron stars. The QCD partition function
is then given in the grand-canonical ensemble as
\beq
{\cal Z}_{QCD}(V,T,\mu_q)={\rm Tr}\exp{\{-(\hat H_{QCD}-\mu_q\hat N_q)/T\} }
\quad ,
\eeq
where $\hat H_{QCD}$ denotes the QCD Hamiltonian, $\hat N_q$ the quark
number operator and $\mu_q$ the quark chemical potential. Statistical
expectation of operators are then given as
\beq
\tave{\hat O}={\cal Z}^{-1}\sum_n\bra{n}\hat O\ket{n}\exp{\{-(E_n-\mu_q N_n)/T\} }
\quad ,
\eeq
where $E_n$ are the exact QCD energies (hadrons). The quark condensate
$\tave{\bar qq}$ can be obtained directly from the free energy density
\beq
\Omega_{QCD}(T,V)=-\lim_{V\to\infty}{T\over V}\ln{\cal Z}_{QCD}(V,T,\mu_q)
\eeq
via the Feynman-Hellmann theorem as
\beq
\tave{\bar qq}={\partial\Omega_{QCD}\over\partial m^\circ_q}
\quad ,
\eeq
where $m^\circ_q$ denotes the bare (current) quark mass. An obvious first
step is to approximate the free energy density by an ideal gas of hadrons.
Using the Gell-Mann Oakes Renner relation for the vacuum chiral condensate
\beq
m_\pi^2f_\pi^2=-2m^\circ_q\ave{\bar qq} \quad ,
\eeq
where $m_\pi$ is the pion mass and $f_\pi$ the pion weak-decay constant,
one then finds 
\beq
{\tave{\bar qq}\over\ave{\bar qq}}=1-
\sum_h{\Sigma_h\varrho^s_h(\mu_q,T)\over f_\pi^2 m_\pi^2} \quad.
\label{qqt}
\eeq
Here
\beq
\Sigma_h=m^\circ_q{\partial m_h\over \partial m^\circ_h}
\eeq
denotes the 'sigma' commutator' (related to the scalar
density of quarks in a given hadron) and $m_h$ the vacuum mass
of a given hadron. At low temperature
and small baryochemical potential ($\mu_B=3\mu_q$), in which 
the hadron gas is
dominated by thermally excited pions and a free Fermi gas of
nucleons,  (\ref{qqt}) leads to the celebrated leading-order result  
\beq
{\tave{\bar qq}\over\ave{\bar qq}}=1-{T^2\over 8f_\pi^2}
-0.3{\rho\over\rho_0} \dots
\eeq
where $\rho_0=0.16$/fm$^3$ is the saturation density of 
symmetric nuclear matter.
This result is model independent and indicates that the mere presence
of an ideal gas of hadrons already alters the vacuum structure and
leads to a decrease of the condensate, without changing the vacuum
properties of the hadrons! Obviously medium-modifications of hadrons
and the corresponding non-trivial change of the QCD vacuum has to
involve hadronic interactions. They become increasingly important as
the medium grows hotter and denser, i.e.\ as the point
of chiral restoration is approached. Thus, the theoretical description
in terms of hadrons
becomes very complex, involving more and more degrees of freedom.
In the vicinity of the restoration transition, in addition, non-perturbative
methods are called for which is far from trivial in effective field theories.

In terms of effective fields $\phi_j(x)$ representing the pertinent 
quark currents $J_{\Gamma_j}(x)$, hadronic correlation 
functions in a hot and dense environment are defined as the 
(retarded) current-current correlation functions
\beq
D_{\phi_j}(\omega,\vec q)=-i\int d^4x\theta(x_0)\tave{[\phi_j(x),\phi_j(0)]}.
\label{prop}
\eeq 
Note that, in contrast to the vacuum, these correlators depend on energy $\omega$
and three-momentum $\vec q$ separately since Lorentz invariance is explicitly
broken by the presence of matter. Equation (\ref{prop}) can be rewritten in terms of
the self energy $\Sigma_{\phi_j}$ as
\beq
D_{\phi_j}(\omega,\vec q)=(\omega^2-\vec q^2-
m_{\phi_j}^2-\Sigma_{\phi_j}(\omega,\vec q))^{-1}
\eeq
where $m_{\phi_j}$ denotes the (bare) mass of the field $\phi_j$ and all
interaction effects are incorporated via $\Sigma_{\phi_j}$ which depends on 
$T$ and $\mu_B$. Given the effective Lagrangian, the objective is then to 
evaluate $\Sigma_{\phi_j}$ as realistic as possible. This is usually attempted
by employing 'chiral counting rules' for evaluating loop 
diagrams contributing to $\Sigma_{\phi_j}$ in the low-density and low-temperature
limit and by adjusting the parameters of the effective Lagrangian to as many
data for elementary scattering processes as available. If everything is done 
consistently, chiral symmetry is properly incorporated and the relation between 
hadronic medium modifications and vacuum changes can be inferred. 

In the vacuum, the hadronic correlators $D_{\phi_j}$ are usually dominated by
a few fairly sharp hadronic 'resonances'. These are visible as 'peaks' in
the corresponding 'spectral functions'
\beq
\rho_{\phi_j}(\omega,\vec q)=-{1\over\pi}{\rm Im}D_{\phi_j}(\omega,\vec q)
\eeq
at the hadronic vacuum mass $m_{\phi_j}$. Two things will happen in the
interacting medium. One the one hand the mass will change to an effective 'pole-mass' 
\beq
m_{\phi_j}^{*2}=m_{\phi_j}^2+{\rm Re}\Sigma_{\phi_j}(m_{\phi_j}^{*2},0) \quad .
\eeq
In fact, one may have \emph{several} solutions and the in-medium
spectrum shows more than one 'peak'. As we shall see, this is usually
the case. On the other hand, the imaginary part ${\rm Im}\Sigma_{\phi_j}$
acquires additional pieces through the interactions with the medium
giving rise to an increased width. If the width becomes too large, the
peak structure is washed out and the notion of a 'quasiparticle'
is lost. This happens when the 'quasiparticle' energy 
\beq
\omega_{\phi_j}^2(\vec q)=\vec q^2+m_{\phi_j}^2+{\rm Re}
\Sigma_{\phi_j}(\omega_{\phi_j}^2,\vec q)
\eeq
is no longer large compared to the width
\beq 
\Gamma_{\phi_j}(\omega_{\phi_j}(\vec q))=
-{1\over 2\pi}{\rm Im}\Sigma(\omega_{\phi_j},\vec q) \quad.
\eeq
This has to be kept in mind when describing in-medium hadrons.

In the following, we will discuss the properties of kaons in the medium
and its consequences for the properties of neutron stars. In
sect.~\ref{sec:AR}, the elementary kaon-nucleon interaction and the
in-medium changes of the kaons as seen in kaonic atoms are described. Chiral
effective interactions are constructed which can describe these data and are used
to extract the optical potential of kaons at finite
density. Section~\ref{sec:JSB} utilizes the results found for the kaon optical
potential to apply it to the equation of state for neutron stars. Implications
for a first order phase transition to a kaon condensed state are
listed. Effects on the onset of kaon condensation by the presence of
hyperons in matter are studied. Finally, we summarize in
sect.~\ref{sec:summary} and give an outlook.


\section{Kaons in dense matter}
\label{sec:AR}

The properties of kaons and antikaons in the nuclear medium
have been the object of numerous investigations since the
possibility of the existence of a kaon
condensed phase in dense nuclear matter was pointed out
by Kaplan and Nelson \cite{KN86}.
If the $K^-$ meson develops sufficient attraction in dense matter
it could
be energetically more favorable,
after a certain critical density, to
neutralize the positive charge with antikaons rather than with
electrons. A
condensed  kaon phase would then start to develop, changing
drastically the
properties of dense neutron star matter
\cite{Brown92,Thorsson94,Fujii96,Li97,Knorren95a,Knorren95b,SM96,GS98L,GS99,TY98}. 
In fact, kaonic atom data, a compilation of which is given in 
\cite{FGB94},
favor an attractive $K^-$ nucleus interaction.
On the other hand, the
enhancement
of the $K^-$ yield in Ni+Ni collisions measured recently by the
KaoS
collaboration at
GSI \cite{kaos97} can be explained by a strong
attraction in the medium for the K$^-$
\cite{cassing97,LKF94,LK96,LLB97}. However, 
the antikaons might feel 
a repulsive potential at the relatively
high temperatures attained in heavy-ion reactions, 
and an alternative mechanism, based on the
production of antikaons via an in-medium enhanced $\pi\Sigma\to K^- p$
reaction, has been suggested \cite{SKE99}. 

The theoretical investigations that go beyond pure phenomenology
\cite{sibir98}
have mainly followed two different
strategies. One
line of approach is that of the mean field models, built within
the
framework of chiral Lagrangians
\cite{LLB97,CHL95,CHL96b,Mao99} or based on the relativistic Walecka
model which are 
extended to incorporate strangeness in the form of hyperons or
kaons 
\cite{Sch97} or by using explicitly the
quark degrees of freedom \cite{Tsushi98}.
The other type of approach aims at obtaining
the in-medium
$\bar{K}N$ interaction microscopically by incorporating the
medium
modifications in 
a $\bar{K} N$ amplitude using (chiral) effective interactions that reproduces the
low energy scattering data 
and generates the $\Lambda(1405)$
resonance dynamically
\cite{alberg76,star87,Koch94,WKW96,Waas97,Lutz98,RO99}.
In this section, we focus on this latter perspective, since it allows
one to
systematically study the importance of all the different mechanisms that
might modify the
${\bar K}N$ interaction in the medium from that in free space.

\subsection{${\bar K}N$ interaction in the medium}

\begin{figure}[htb] 
\begin{center}
\includegraphics[width=\linewidth]{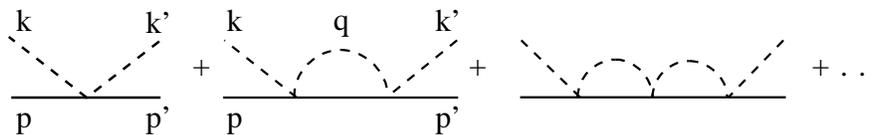}
\end{center}
\caption{
Diagrammatic representation of the Bethe-Salpeter equation for 
${\bar K}N$ scattering.}
\label{fig:fig1kn}
\end{figure}

The free ${\bar K}N$ scattering observables (${\bar K}=K^-$ or $\bar{K}^0$)
are derived from the scattering amplitude, obtained from the Bethe-Salpeter equation
\begin{equation}
T = V + V G T \ ,
\label{eq:BS}
\end{equation}
which is depicted diagrammatically in Fig.~\ref{fig:fig1kn}.
Note that, in the case of ${\bar K}^- p$ scattering, this is a
coupled-channel equation involving ten intermediate states, namely
$K^- p$, $\bar{K}^0 n$, $\pi^0 \Lambda$, $\pi^+ \Sigma^-$, $\pi^0 \Sigma^0$, 
$\pi^-\Sigma^+$, 
$\eta\Lambda$, $\eta\Sigma^0$, $K^+\Xi^-$ and $K^0 \Xi^0$.
The loop operator $G$ stands for the diagonal intermediate meson-baryon
($MB$)
propagator and $V$ is a suitable $MB \to M^\prime B^\prime$ transition
potential.

A connection with the chiral Lagrangian was established in
\cite{NK}, where the
properties of of the $S = -1$ meson-baryon sector were studied in a
potential model \cite{NK},  such that, in Born approximation,
it had the same S-wave scattering length as the chiral Lagrangian,
including both the lowest order and the momentum dependent $p^2$ terms. 
No $\eta\Lambda$, $\eta\Sigma$ or $K\Xi$
channels
were considered,
on the basis that they were not
opened at the $K^- p$ threshold.
By fitting five parameters,
corresponding to, so far, unknown parameters of the second order
chiral Lagrangian plus the range parameters of the potential, the
$\Lambda (1405)$ resonance was generated as a quasi-bound meson-baryon state
and the $K^- p \rightarrow M B$ cross sections, as well as
the available branching rations at threshold, were well reproduced.
Note that being close to a resonance 
forces one to resort to non-perturbative approaches, such as
summing the infinite Bethe-Salpeter
series as represented diagrammatically in Fig.~\ref{fig:fig1kn}.
A recent work \cite{AO}, which shares many
points with \cite{NK}, showed that
all the strangeness $S=-1$ meson-baryon scattering observables  near
threshold could be reproduced
using only the lowest order Lagrangian in the Bethe-Salpeter equation and one
parameter, the
cut-off $q_{\rm max}$ that regularizes the loop function $G$.  
All 10 meson-baryon
states that can be generated from the octet of pseudoscalar
mesons and the
octet of ground-state baryons were included, the additional $\eta
\Lambda$ state being
a quite relevant one. 
The success of this method in reproducing the scattering observables 
is analogous to that
obtained in the meson-meson sector \cite{oller97}. An explanation
was found in \cite{oller99} by applying the
Inverse Amplitude Method in coupled channels to the same problem,
with the lowest and next-to-lowest order meson-meson
Lagrangian. It was shown that
the selection of an appropriate cut off for a particular $I,J$
channel could minimize the contribution of the next-to-lowest order
Lagrangian, reducing the relevant higher order terms to those
iterated
by the Bethe-Salpeter equation with the lowest order Lagrangian, which is
the simplified method followed 
in \cite{AO,oller97} (see also the discussion in the review \cite{report}).

The most obvious medium effect is that induced by 
Pauli-blocking on the nucleons in the
intermediate states.
This makes the
${\bar K}N$ interaction density dependent and modifies, in turn,
the $K^-$
properties from those in free space. These effects
were already included a long time ago 
in the context of Brueckner-type many body theory, 
using a separable $\bar{K}N$ interaction \cite{alberg76}. Some more recent
works \cite{Koch94,WKW96,Waas97,Lutz98,RO99} take the
$\bar{K}N$
interaction from the chiral Lagrangian in S-wave.
However,
as seen in the
recent work by Tolos et al.~\cite{laura} which uses the Bonn ${\bar K}N$
potential model \cite{bonn}, 
the incorporation of higher angular momenta has 
non-negligible effects on the properties of the 
antikaon at high momentum, which is of relevance for the analysis of in-medium
effects for the K$^-$ in heavy-ion collisions.

In the actual calculations, the effect of Pauli blocking is
 incorporated by replacing the free nucleon propagator in the 
loop function $G$ for intermediate ${\bar K}N$ states
by an in-medium one of the type
\begin{equation}
A(\sqrt{s}-q^0,-\vec{q},\rho)=
\frac{1-n(\vec{q}_{\rm
lab})}{\sqrt{s}-q^0-E(-\vec{q}\,)+i\epsilon} +
\frac{n(\vec{q}_{\rm
lab})}{\sqrt{s}-q^0-E(-\vec{q}\,)-i\epsilon} \ ,
\label{eq:nucleon}
\end{equation}
where $n(\vec{q}_{\rm lab})$ is the occupation
probability of a
nucleon of
momentum $\vec{q}_{\rm lab}$ in the lab frame.

The most spectacular consequence of the Pauli principle is that the
blocking of
intermediate states shifts the resonance to
higher energy.
This changes the
${\bar K}N$ interaction at threshold from being repulsive in free
space to
being attractive in the medium.
Therefore, antikaons develop an attractive optical potential which,
incorporated again in the ${\bar K}N$ states of the in-medium
scattering equation, may compensate the upward shifting of these
intermediate states
induced by Pauli blocking. This feedback has been recently
confirmed by the calculation
of Lutz
\cite{Lutz98}, where the dressing of the antikaon is incorporated in the 
intermediate loops in a self-consistent manner.
The $\Lambda(1405)$ resonance
remains then unchanged, in qualitative agreement with what was
noted in \cite{star87} using a constant mean field potential for
the $\bar{K}$.

Since the dressing of the antikaon turns out to be so relevant, one might
wonder about dressing the other mesons or baryons that play a role in the
${\bar K}N$ system. This has been explored in the recent work \cite{RO99},
where a self-consistent 
antikaon self-energy is obtained including the dressing of 
the pions in the $\pi \Lambda$, $\pi \Sigma$ intermediate
states, which are the ones that couple
strongly to ${\bar K} N$.

Incorporating the medium modified mesons in the calculation is
technically achieved by replacing the free meson propagator in the loop 
function $G$ by
\begin{equation}
D(q^0,\vec{q},\rho) = \frac{1}{(q^0)^2-{\vec q\,}^2 - m^2 -
\Pi(q^0,\vec{q},\rho)} =
\int_0^\infty d\omega \, 2\omega\,
\frac{S(\omega,\vec{q},\rho)}{(q^0)^2-\omega^2 + i\epsilon} \ ,
\label{eq:meson}
\end{equation}
where $\Pi(q^0,\vec{q},\rho)$ is the meson self-energy. The Lehman
representation shown in the  
second equality of (\ref{eq:meson}) introduces the 
spectral density,
$S(\omega,\vec{q},\rho)=-{\rm Im}
D(\omega,\vec{q},\rho)/\pi$, which in the case of free mesons
reduces to
$\delta(\omega-\omega(\vec{q}\,))/2\omega(\vec{q}\,)$.
With these modifications the loop
integral becomes
\begin{eqnarray}
G(P^0,\vec{P},\rho) &=&
\int_{\mid\vec{q}\,\mid < q_{\rm max} } \frac{d^3 q}{(2 \pi)^3}
\frac{M}{E (-\vec{q}\,)}
\int_0^\infty d\omega \,
 S(\omega,{\vec q},\rho) \nonumber \\
&\times & \left\{
\frac{1-n(\vec{q}_{\rm lab})}{\sqrt{s}- \omega
- E (-\vec{q}\,)
+ i \epsilon} +
\frac{n(\vec{q}_{\rm lab})}
{\sqrt{s} + \omega - E(-\vec{q}\,) - i \epsilon } \right\} \ ,
\label{eq:gmed}
\end{eqnarray}
where $(P^0,\vec{P})$ is the total four-momentum in the lab frame
and
$s=(P^0)^2-\vec{P}^2$. Note that the loop function $G$ does-not contain
the $V$ and $T$ amplitudes. This simplification is possible in the 
chiral approach 
of \cite{AO}, where it was shown that the 
the amplitudes
factorize on-shell out of the loop, since the
off shell part could be absorbed into a renormalization of the
coupling constant $f_\pi$. 
These arguments can not be applied
in potential models
and a coupled system of integral equations must be solved.
A reasonable simplification is obtained if the dressing of the antikaon in
the loop function $G$ is taken into account via an energy-independent
self-energy,
$\Pi(q^0=\varepsilon_{qp}(\vec{q}\,),\vec{q},\rho)$,
evaluated at the quasiparticle energy, $\varepsilon_{qp}(\vec{q}\,)$,
which
is the solution of the in-medium dispersion relation 
\begin{equation}
\varepsilon^2_{qp}(\vec{q}\,)=\vec{q}\,^2 + m_K^2 +
\Pi(q^0=\varepsilon_{qp}(\vec{q}\,),\vec{q},\rho)  \ .
\label{eq:qp}
\end{equation}
This is the prescription followed in \cite{SKE99,laura} and amounts 
to approximate the actual antikaon spectral function by a symmetric
pseudo-Lorentzian peak at the quasiparticle energy given by 
(\ref{eq:qp}). With this assumption, the
loop function $G$ looks like the free
one, but replacing the free antikaon energy $\sqrt{\vec{q}\,^2 + m_K^2}$
by the complex quasiparticle one, $\varepsilon_{qp}(\vec{q}\,)$.

One might also include the dressing of the baryons by
assuming that the single particle energy
$E(-\vec{q}\,)$ in (\ref{eq:nucleon}) to contain a mean-field
potential of
the type $U_0 \times \rho/\rho_0$, 
as done in \cite{RO99,laura}. For the
nucleon, a
reasonable depth value
is
$U_0^N=-70$ MeV, as suggested by numerous calculations of the
nucleon 
potential in nuclear matter. For the $\Lambda$ hyperon, 
one can
take $U_0^\Lambda=-30$ MeV, as implied by extrapolating 
the experimental
$\Lambda$ single
particle energies in $\Lambda$ hypernuclei to bulk matter
\cite{moto90}. For the 
$\Sigma$ hyperon, there is no conclusive information on the
potential. Early phenomenological analyzes \cite{Batty78}
and calculations
\cite{oset90}  
found the $\Sigma$ atomic data to be compatible with $U_0^\Sigma
\sim -30$
MeV, but more recent analysis indicate a repulsive
potential in the nuclear interior \cite{Batty94}. 
As shown in \cite{RO99}, changing the $\Sigma$ depth
from $-30$ to $+30$ MeV does not change
the results for the properties of the K$^-$ in the medium considerably.

\begin{figure}[htb] 
\begin{center}
\includegraphics[width=0.8\linewidth]{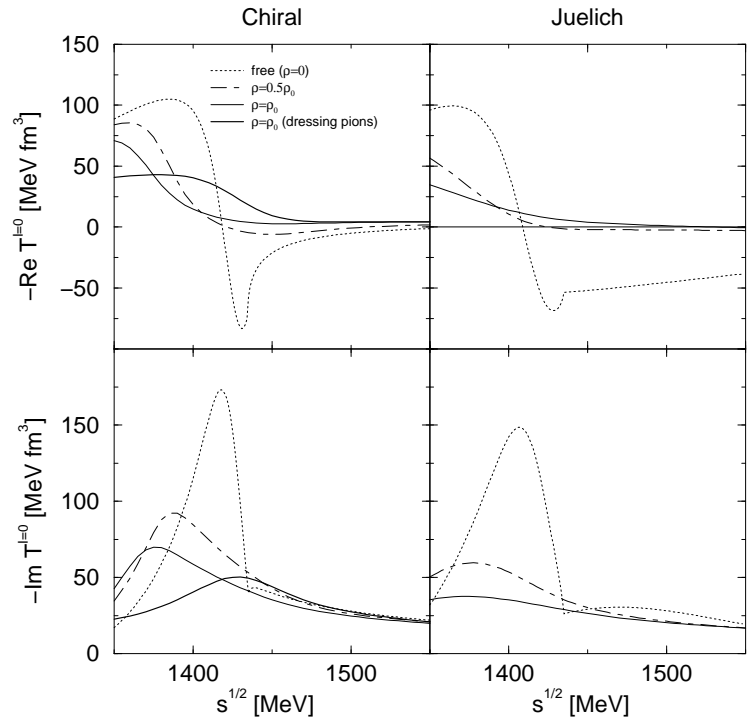}
\end{center}
\caption{Real and imaginary parts of the $I=0$ ${\bar K}N$ scattering amplitude as
functions of $\sqrt{s}$ for $\mid \vec{p}_K + \vec{p}_N\mid = 0$ and
several densities (from \cite{RO99} and \cite{laura}).}
\label{fig:ampl}
\end{figure}

Using the dressed
meson-baryon loop
in the coupled-channel Bethe-Salpeter
equation, one obtains 
the in-medium $\bar{K}N$ amplitude $T_{\rm eff}(P^0,\vec{P},\rho)$.
In Fig.~\ref{fig:ampl} we compare the free amplitude
($\rho=0$, dotted line) in the $I=0$, $L=0$
channel with that obtained at nuclear densities $\rho=\rho_0=0.17$
fm$^{-1}$ (solid line)
and $\rho=0.5\rho_0$ (dot-dashed line).
Two different models are shown,
that of \cite{RO99}, based on the lowest order meson-baryon chiral
Lagrangian, and that of \cite{laura}, based on the potential model
of
the Bonn group \cite{bonn}. In spite of the appreciable differences seen
in the
free scattering amplitudes, the medium modified ones show the same
qualitative trends. Note also how the real part of the amplitude 
(upper panels) at the $K^- p$ threshold
($\sqrt{s}=1433$ MeV) is repulsive in free space and attractive in the
medium. The thick solid line on the left panels show the effect 
on the $K^- p$ amplitude by
dressing the pions in the intermediate states \cite{RO99}.

Most of the available models study the in-medium ${\bar K}N$
amplitude in S-wave.
However, if one aims at extracting the properties of antikaons
through the
analysis of heavy-ion collisions, one must keep in mind that they are
created at a finite momentum of around $250-500$ MeV/c, hence the effect
of higher partial waves might be relevant. The meson-exchange
${\bar K}N$ potential of the
J\"ulich group \cite{bonn} is given in partial waves and the
main results from a recent  study \cite{laura} on the
effect of the angular momentum states higher than the commonly 
considered $L=0$ one will
be summarized in the next
subsection. From the chiral perspective, the P-wave amplitudes from the
next-to-leading order ${\bar K}N$ chiral Lagrangian have been identified
and, recently, the parameters have been fitted to reproduce a large amount
of low energy data \cite{caro}.
However, a nuclear medium application of this model is not available yet.

\subsection{In medium ${\bar K}$ properties}

The $\bar{K}$ self-energy
is obtained by summing the
in-medium $\bar{K}N$ interaction, $T_{\rm
eff}(P^0,\vec{P},\rho)$,
over the nucleons in the Fermi sea
\begin{equation}
\Pi_{\bar{K}}(q^0,{\vec q},\rho)=2\sum_{N=n,p}\int
\frac{d^3p}{(2\pi)^3}
n(\vec{p}\,) \,   T_{\rm eff}(q^0+E(\vec{p}\,),\vec{q}+\vec{p},\rho) \ .
\label{eq:selfka}
\end{equation}
Note that a self-consistent approach is required since one
calculates the ${\bar K}$ self-energy from the effective
interaction $T_{\rm eff}$ which uses ${\bar K}$ propagators which
themselves include the self-energy being calculated.

\begin{figure}[htb] 
\begin{center}
\includegraphics[width=0.7\linewidth]{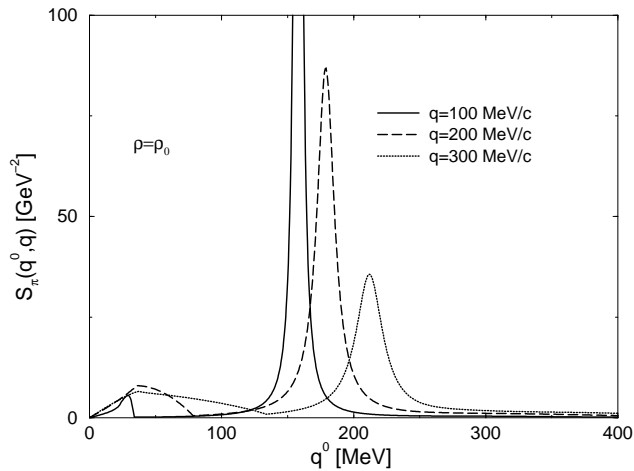}
\end{center}
\caption{Pion spectral density at $\rho=\rho_0$ for several
momenta (from \cite{RO99}).}
\label{fig:specpi}
\end{figure}

A P-wave contribution to the ${\bar K}$
self-energy coming from the coupling of the ${\bar K}$ meson to
hyperon-hole excitations can be easily included 
(if it is not already contained in $T_{\rm
eff}$) and the expression can
be found in \cite{RO99}.
In that work the pions are also dressed
through a pion self-energy that contains  one- and two-nucleon absorption
and is conveniently
modified to include the effect of nuclear short-range
correlations (see \cite{ramos94} for details).
The resulting pion spectral density 
in nuclear matter at density $\rho=\rho_0$ is shown in
Fig.~\ref{fig:specpi} 
as a function of the pion energy for several
momenta.  The strength is distributed over a wide range of
energies and,
as the pion momentum increases, the
position of the peak is
increasingly lowered from the corresponding one in free space as
a consequence of the attractive pion-nuclear potential.
Note that, to the left of the peaks, there appears the typical
structure
of the $1p1h$ excitations which give rise to $1p1h\Lambda$ and
$1p1h\Sigma$ components in the effective ${\bar K}N$ interaction.
Although not visible in the linear scale used in 
Fig.~\ref{fig:specpi}, there is some additional strength at energies
around 300 MeV associated to the excitation of the $\Delta$ resonance.

\begin{figure}[htb]
\begin{center}
\includegraphics[width=0.45\linewidth]{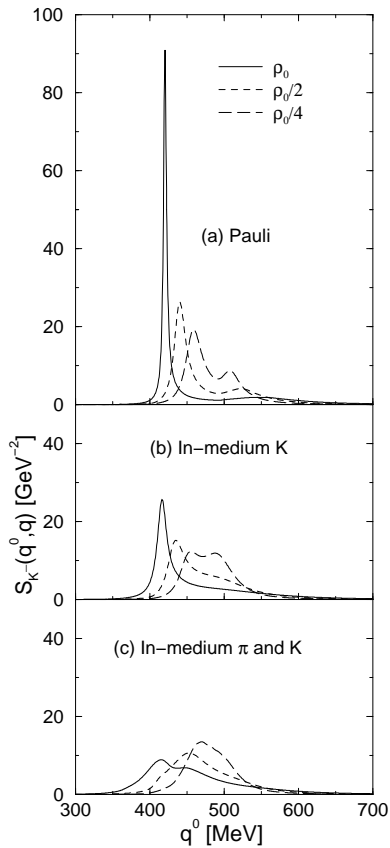}
\end{center}
\caption{$K^-$ spectral density for zero momentum from the
chiral model of ref. \cite{RO99}}
\label{fig:kspec}
\end{figure}

The spectral function of a $K^-$ meson of zero
momentum obtained with the chiral model of \cite{RO99} is
shown in Fig.~\ref{fig:kspec} for various densities: $\rho_0$, $\rho_0/2$
and
$\rho_0/4$.
The results in the upper panel include
only Pauli blocking effects, i.e. the nucleons propagate as in
(\ref{eq:nucleon}) but the mesons behave as in free space. 
At $\rho_0/4$ one clearly sees two excitation modes. The left one
corresponds to the
$K^-$ pole branch, appearing
at an energy smaller than the kaon mass, $m_K$, due to the
attractive medium effects. The peak on the right corresponds to
the
$\Lambda(1405)$-hole excitation mode,
located above $m_K$ because of the shifting of
the $\Lambda(1405)$ resonance to energies above the $K^-p$
threshold.
As density increases, the $K^-$ feels an enhanced
attraction while
the $\Lambda(1405)$-hole peak moves to higher energies and
loses
strength, a reflection of the tendency of the $\Lambda(1405)$
to dissolve in the dense nuclear medium. These features were
already
observed in \cite{Koch94,Waas97}.
The (self-consistent) incorporation of the ${\bar K}$ propagator
in
the Bethe-Salpeter equation softens the effective interaction,
$T_{\rm eff}$, which becomes
more spread out in energies (solid and dot-dashed lines in
Fig.~\ref{fig:ampl}). The resulting $K^-$ spectral
function (middle
panel in Fig.~\ref{fig:kspec}) shows the displacement of the resonance
to lower energies
because, as already noted, the attraction felt by
the ${\bar
K}$ meson lowers the threshold for the
${\bar K}N$ states that had been increased by the Pauli blocking
on the
nucleons. 
This has a compensating effect
and the resonance moves backwards, slightly below its free
space value.
The $K^-$ pole peak appears at similar or slightly
smaller energies, but its width is larger, due to the
strength of the intermediate ${\bar K}N$ states being distributed
over a wider region of energies.
Therefore the $K^-$ pole and the $\Lambda(1405)$-hole branches
merge into one another and can hardly be distinguished.
Finally, when the pion is dressed according to the spectral
function
shown in Fig.~\ref{fig:specpi} the effective interaction
$T_{\rm eff}$ becomes even smoother (thick solid lines in
Fig.~\ref{fig:ampl}). The resulting 
$K^-$ spectral function is displayed in the bottom panel in
Fig.~\ref{fig:kspec}. 
Even at
very small densities one can no longer distinguish the
$\Lambda(1405)$-hole
peak from the $K^-$ pole one.
As density increases, the attraction
felt by the $K^-$ is more moderate and the $K^-$ pole peak
appears at
higher energies than in the other two approaches.
However, more strength is found at very
low energies, especially at $\rho_0$, due to the $1p1h$ $2p2h$ components
of the pionic strength, which couple the
${\bar K}N$ state to the
$1p1h\Sigma$ and $2p2h\Sigma$ ones.
It is precisely the opening of the
$\pi\Sigma$ channel, on top of the already opened $1p1h\Sigma$
and
$2p2h\Sigma$ ones, which causes a cusp structure to appear
slightly above 400 MeV.

\begin{figure}[htb]
\begin{center}
\includegraphics[width=0.55\linewidth,angle=-90]{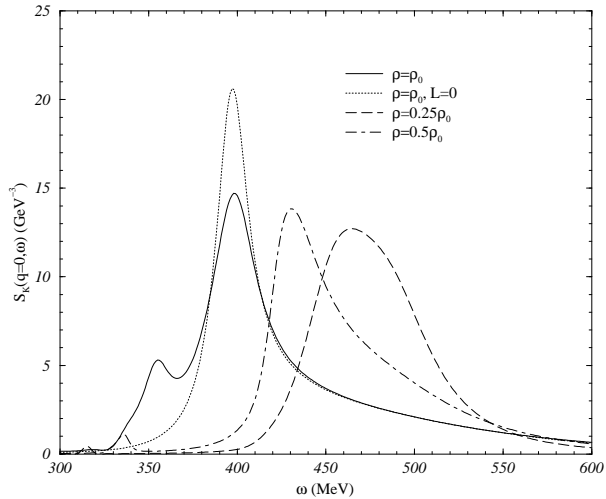}
\end{center}
\caption{$K^-$ spectral density for zero momentum 
  using the Bonn ${\bar K}N$ potential (from \cite{laura}).
 }
\label{fig:kspec2}
\end{figure}

The calculation of \cite{laura} using the
Bonn ${\bar K}N$ potential obtains similar results, which are
shown in Fig.~\ref{fig:kspec2} for
the same three densities.
We notice some structure of the spectral
function to the left of the quasiparticle peak at energies
of the ${\bar K}$ around $320-360$ MeV. This is the 
in-medium reflection of a singularity in the
$L=1$, $I=1$ free space amplitude around the mass of the $\Sigma$ baryon
induced by the $\Sigma$-pole diagram present in the Bonn ${\bar K}N$
potential \cite{bonn}.
This peak is therefore indicating the physical excitation of
$\Sigma h$ states with antikaon quantum numbers.
The dotted line shows the spectral density at $\rho=\rho_0$ but keeping
only the $L=0$ component of the ${\bar K}N$ interaction. In agreement
with the behavior of the complex antikaon potential at zero momentum shown
below, we
observe
that the location of the quasiparticle peak (driven by the real part) only
moves a few MeV, while
the width (driven by the imaginary part) gets reduced by about
30\% when including higher partial waves.

One may define a non-relativistic  antikaon single-particle potential from
the self-energy
at the quasiparticle energy via the relation
\begin{equation}
U_K(q)=
\frac{\Pi_{\bar{K}}(\varepsilon_{qp}(\vec{q}\,),{\vec q},\rho)}{2 m_K} \ .
\label{eq:upot}
\end{equation}

\begin{figure}[htb]
\begin{minipage}{.48\linewidth}
\centerline{
     \includegraphics[width=\textwidth]{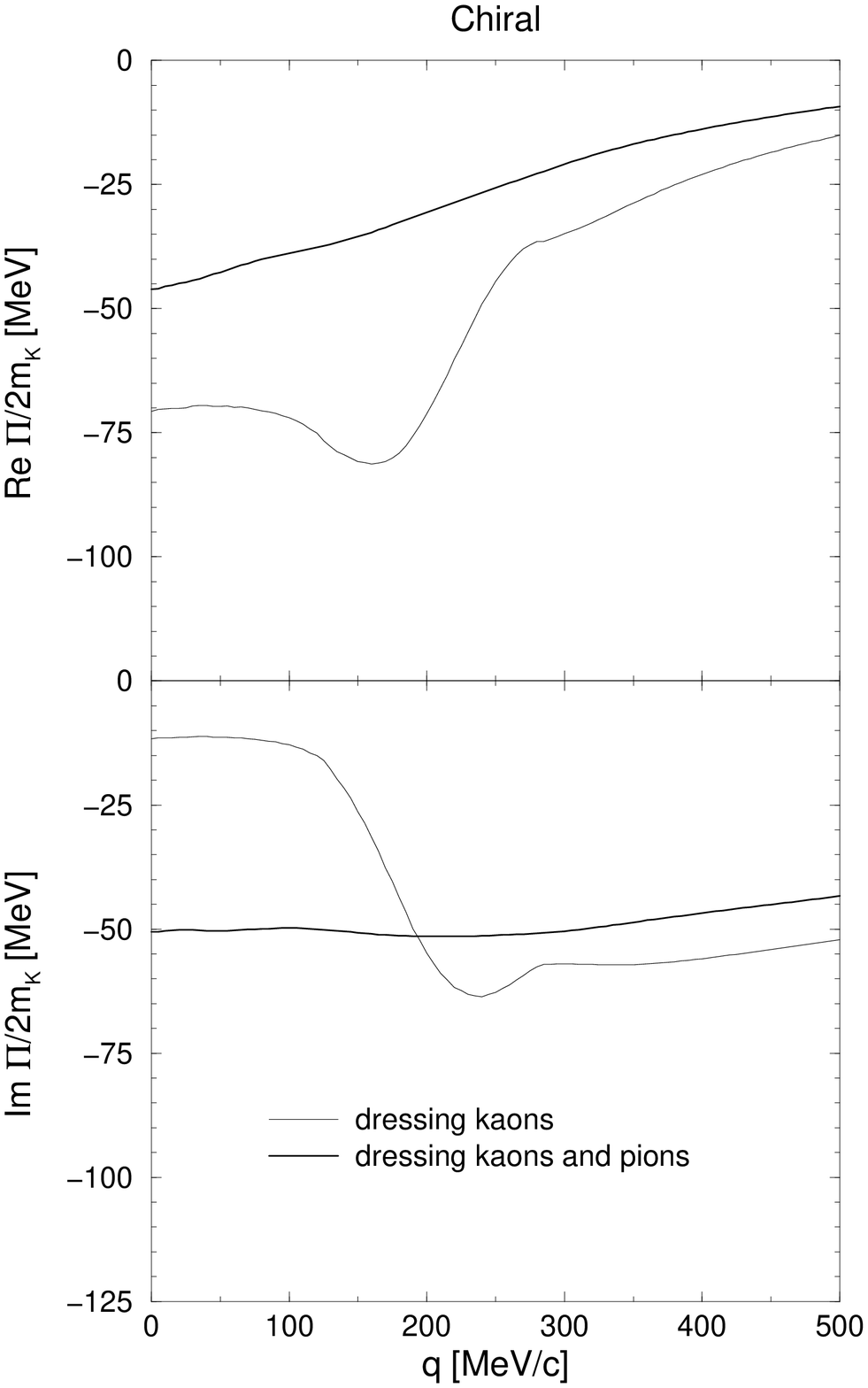}
}
      \caption{\small
Real and imaginary parts of the ${\bar K}$ optical potential
at $\rho=\rho_0$ as functions of the antikaon momentum, as obtained from
the chiral model of \cite{RO99}.}
        \label{fig:upot1}
\end{minipage}
\hfill
\begin{minipage}{.48\linewidth}
\centerline{
     \includegraphics[width=\textwidth]{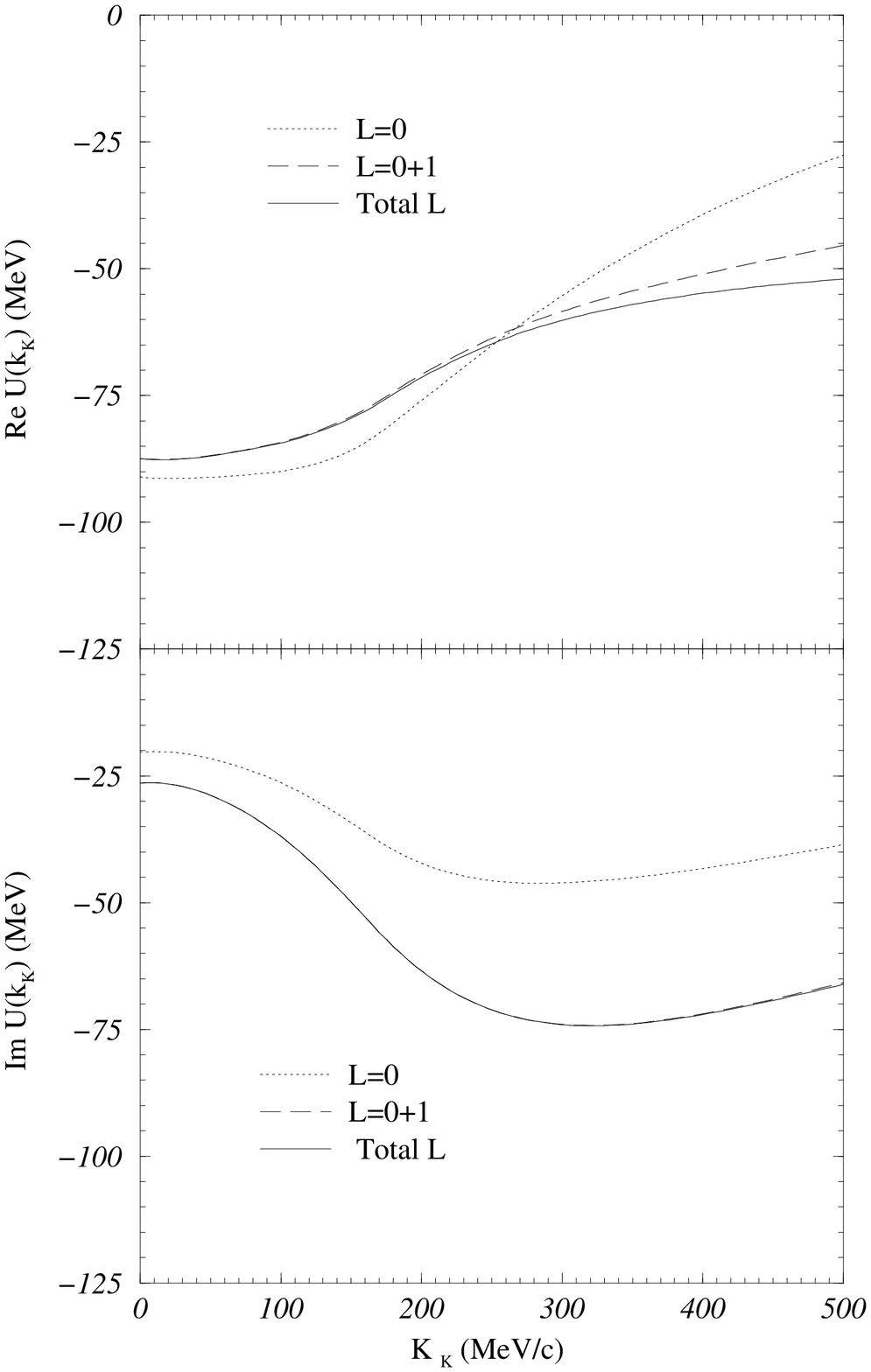}
}
      \caption{\small
The same as Fig.~\ref{fig:upot1}, but 
obtained using the Bonn ${\bar K}N$ potential
(taken from \cite{laura}).}
        \label{fig:upot2}
\end{minipage}
\end{figure}

The real and imaginary parts of the antikaon potential at $\rho=\rho_0$,
obtained from the chiral model of ref.~\cite{RO99}, are shown in
Fig.~\ref{fig:upot1} as function of the 
antikaon momentum for two approximations, one in which only 
the antikaon self-energy is considered in the intermediate loops 
(thin solid lines) and another in which the
pions are also dressed (thick solide line). 
Note that the antikaon potential obtained when
the pions are also dressed has much less structure. This is due to the
smoother in-medium amplitude, but also to the different quasiparticle
energy at which the antikaon self-energy is evaluated. This quasiparticle
energy is more attractive when only antikaons are dressed and, hence,
the amplitude is explored at lower energy regions, closer to the position
of the in-medium $\Lambda(1405)$ resonance.
Dressing the intermediate pions gives an antikaon potential depth at zero
momentum of $-45$ MeV. This is
about half the attraction of that obtained with other
recent models and approximation schemes
\cite{LLB97,Mao99,Sch97,Tsushi98,WKW96,Waas97}, which give rise to
potential depths at the center of the nucleus
between $-140$ and $-75$ MeV.
Results obtained by the model of \cite{laura} using the Bonn ${\bar
K}N$ potential \cite{bonn}, where only
kaons are dressed, are shown in Fig.~\ref{fig:upot2}.
In this figure one can also see the effect
of including the higher partial waves of the Bonn ${\bar K}N$
interaction. We observe that the antikaon nuclear potential at zero
momentum receives some contribution of partial waves
higher than $L=0$, due to the fact that the ${\bar K}$ meson interacts
with nucleons that occupy states up
to the Fermi momentum, giving rise to finite ${\bar K}N$
relative momenta of up to around 90 MeV/c. Clearly, the effect of the
higher partial waves increases with increasing ${\bar K}$ momentum,
flattening out the real part of the potential and producing more
structure to the imaginary part. At an antikaon momentum of around 500
MeV/c,
the inclusion of the higher partial waves practically doubles the
size of the antikaon potential with respect to the S-wave value.

\subsection{Kaonic atoms}

\begin{figure}[htb] 
\begin{center}
\includegraphics[width=0.6\linewidth]{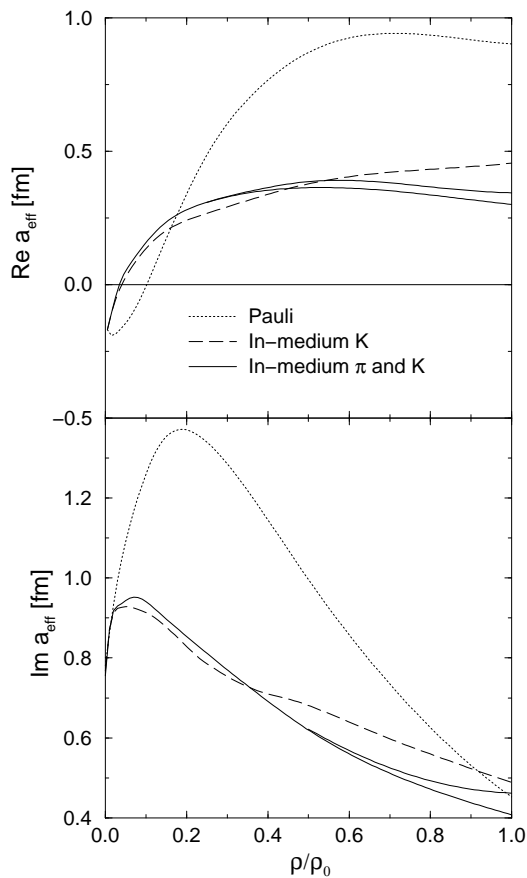}
\end{center}
\caption{$K^-N$
scattering length
as a function of density (from \cite{RO99}).}
\label{fig:scatlen}
\end{figure}

Since
the $K^-$ in kaonic atoms are bound with small (atomic) energies, their
study requires the knowledge of the
antikaon self-energy at 
$(q^0,\vec{q})=(m_K,\vec{0})$. 
The real and imaginary parts of the isospin averaged in-medium
scattering length, defined as
\begin{equation}
a_{\rm eff}(\rho)= -\frac{1}{4\pi} \frac{M}{m_K + M }
\frac{\Pi_{\bar{K}}(m_K,\vec{q}=0,\rho)}{\rho} \ ,
\end{equation}
obtained from the chiral model of \cite{RO99}, are shown
in Fig.~\ref{fig:scatlen}
as function of the nuclear density
$\rho$. The change of the real part of the effective scattering length ${\rm Re}\,
a_{\rm eff}$ from negative to positive values indicates the
transition from a repulsive interaction in free space to an
attractive one in the medium. As shown by the dotted
line, this transition happens at a density of about $\rho\sim
0.1\rho_0$ when only Pauli effects are considered, in agreement
with what was found in \cite{Koch94,WKW96}. However, this
transition occurs at even lower densities ($\rho \sim 0.04
\rho_0$) when one considers the self-energy
of the mesons in the description, whether one dresses only the
$\bar{K}$
meson (dashed line) or both the $\bar{K}$ and $\pi$ mesons (solid
line).
The deviations from the approach including only Pauli
blocking or those dressing the mesons are quite appreciable over
a wide range of densities. The thin solid lines show the results
obtained
with a repulsive $\Sigma$ potential depth of
$+30$ MeV.
The deviations from the thick solid line, obtained for an
attractive
potential depth of $-30$ MeV, are smaller than 10\%
and only show
up at higher densities.

\begin{figure}[ht!] 
\begin{center}
\vspace*{-1cm}
\includegraphics[width=0.7\linewidth]{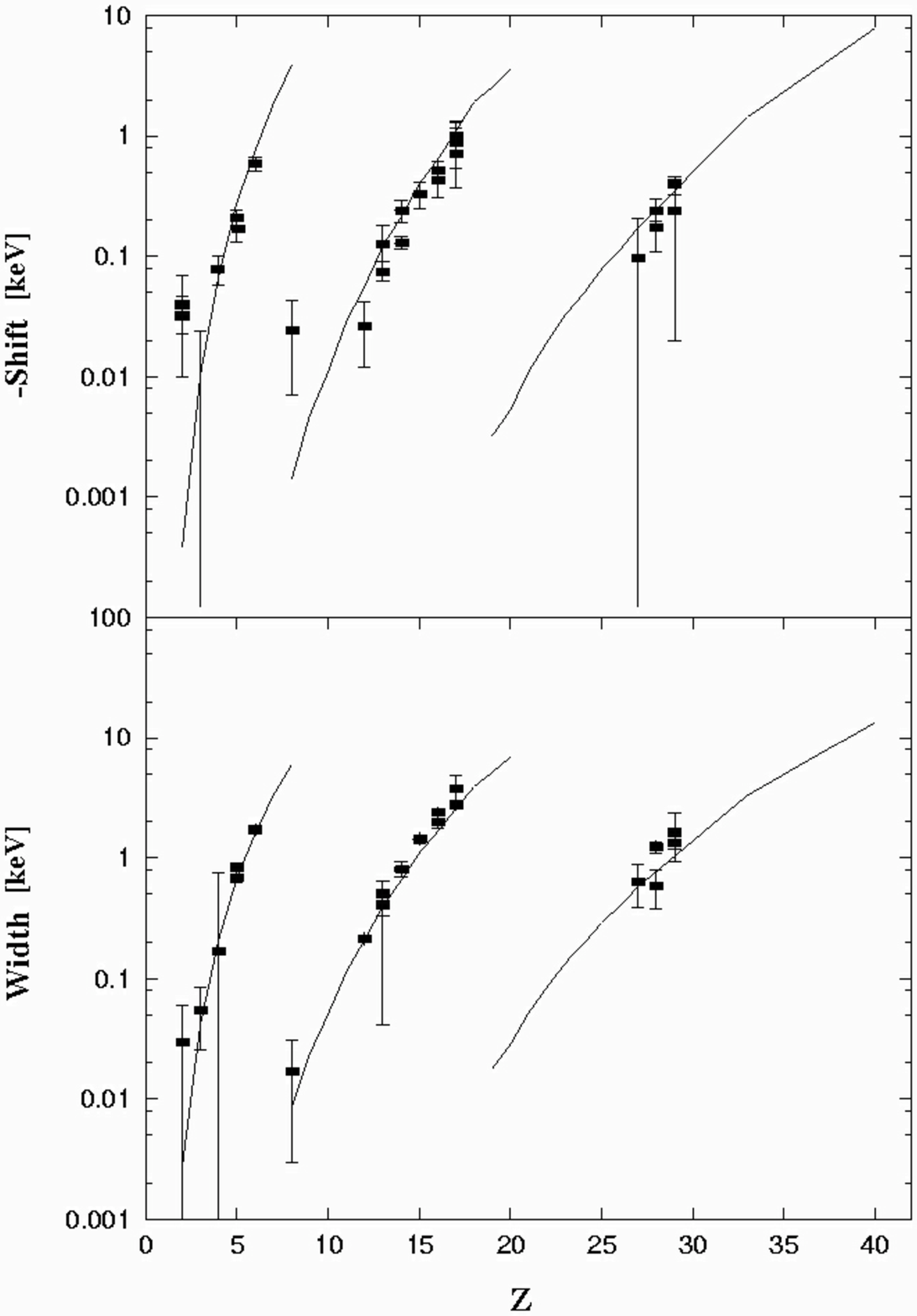}
\end{center}
\caption{Energy shifts and widths of kaonic atom states 
(from 
 \cite{oku99}). The experimental data are taken from
the compilation given in \cite{BGF97}.}
\label{fig:katom}
\end{figure}

The implications of the
density dependent scattering length
displayed in Fig.~\ref{fig:scatlen}
on kaonic atoms have been recently analyzed \cite{oku99}
in the framework of a local density approximation, which amounts to
replace the nuclear matter density $\rho$ 
by the density profile $\rho(r)$ of the particular nucleus. 
The results displayed in Fig.~\ref{fig:katom} show that
both the energy shifts
and widths of kaonic atom states agree well with the
bulk of experimental data \cite{BGF97}.

Reproducing kaonic atom data with this moderately attractive
antikaon nucleus potential of $-45$ MeV is in contrast with
the depth of around $-200$ MeV obtained
from a best fit to $K^-$ atomic data with a phenomenological
potential that includes an additional
non-linear density
dependent term \cite{FGB94}. 
A hybrid model, combining a relativistic
mean field
approach in the nuclear interior and a phenomenological density
dependent potential at the surface that is fitted to $K^-$ atomic
data, also favors a strongly attractive $K^-$ potential of depth
$-180$ MeV \cite{FGMC99}.
On the other hand,
the early Brueckner-type calculations of 
\cite{alberg76} also obtained a shallow $K^-$-nucleus potential,
of
the order of $-40$ MeV at the center of $^{12}$C, and predicted
reasonably well the $K^-$ atomic data available at that time.
Acceptable fits to kaonic atom data have also been obtained using
charge
densities and
a phenomenological $T_{\rm eff} \rho$ type potential
with a depth of the
order of $-50$ MeV in the nuclear interior \cite{batty81}, which goes down
to $-80$ MeV,
when
matter densities are used instead \cite{FGB94}. 

A clarifying quantitative comparison of kaonic atom results obtained
with various $K^-$-nucleus potentials can be found in \cite{baca00}.
There, the reasonable reproduction of data obtained 
with the chiral antikaon-nucleus potential of \cite{RO99}, shown
in Fig.~\ref{fig:katom}, is
quantified with a $\chi^2/d.o.f. = 3.8$. This potential is then modified by
an additional phenomenological
piece, linear in density, which is fitted to the known data and is able to
bring the agreement
to the level of $\chi^2/d.o.f. = 1.6$. This results into a final potential
which is slightly more attractive ($-50$ MeV at $\rho_0$) and has a 
reduced imaginary part  by about a factor 2. The work \cite{baca00} reemphasizes
that kaonic atoms only explore the antikaon potential at the 
surface of the nucleus. Therefore,
although all models predict attraction for
the $K^-$-nucleus
potential, the precise
value of its depth at the center of the nucleus, which has important
implications for the occurrence of kaon condensation, is still not known.
It is then
necessary
to gather more data that could help in disentangling the
properties of
the $\bar{K}$ in the medium. Apart from the valuable information
that can
be extracted from the production of $K^-$
in heavy-ion collisions, one could also measure deeply bound
kaonic states,
which have been predicted to be narrow \cite{oku99,baca00,FG99a} and
could be measured
in $(K^-,\gamma)$ \cite{oku99} or $(K^-,p)$ reactions
\cite{FG99b,Kishi99}.


\section{Kaons in neutron stars}
\label{sec:JSB}

Kaon condensation has first thought to be irrelevant for neutron stars as
their mass has to be lowered so drastically to appear in beta-stable neutron
star matter \cite{Glen85}. Nevertheless, as demonstrated in the last sections, 
the in-medium effects for the kaons, especially for the K$^-$ can be quite
pronounced which reopened the issue of kaon condensation for neutron stars
\cite{KN86}.  The topic has been extensively discussed in the literature over
the last years (see e.g.\
\cite{Brown92,Thorsson94,Fujii96,Li97,Knorren95a,Knorren95b,SM96,GS98L,GS99}).
In all these approaches, the antikaon-nucleon interaction has been parameterized
in effective field theoretical models which were guided by the investigations of the last
sections. So far, a consistent coupled channel calculation incorporating a
realistic nucleon-nucleon interactions as well as kaon-nucleon interactions has not
been performed due to the complexity of the problem. We will outline in this
section, how one can parameterize the antikaon-nucleon interaction in a simple
field theoretical model and apply it to the equation of state (EoS) of
beta-stable matter and neutron stars.

\subsection{Effective model of kaon-nucleon interactions}

In neutron star matter, only baryon number and charge are conserved. Hence,
kaons or antikaons, as well as hyperons, can appear inside neutron
stars by strangeness changing processes. The onset of the appearance of the
negatively charged K$^-$ is given by the equality of the effective antikaon
chemical potential (or effective antikaon energy) in matter with the
electrochemical potential 
\begin{equation}
\omega_K = \mu_{K^-} = \mu_e
\quad .
\label{eq:cond}
\end{equation}
Then processes like
\begin{equation}
e^- \to K^- + \nu_e \qquad n \to p + K^-
\end{equation}
are energetically allowed. The K$^-$ is replacing electrons from the Fermi
surface or equivalently transforming neutrons to protons. The K$^-$ as a boson
will form a condensate with zero momenta as the s-wave interaction with
nucleons is attractive. The presence of the zero momentum K$^-$'s, compared to
the high momenta electrons, will lower the overall energy of the system. Also, 
the increase in the proton fraction will lower the isospin asymmetry of the
matter. As the nuclear asymmetry term is strongly repulsive, K$^-$ can again lower the 
energy of the system substantially.

Guided by the discussion of the previous sections, we write down now an
effective Lagrangian which models the kaon-nucleon interaction:
\begin{equation}
{\cal L}_K = {\cal D}_\mu^* K^* {\cal D}^\mu K - {m^*_K}^2 K^*K
\label{eq:lagK}
\end{equation}
where the vector fields are coupled minimally 
\begin{equation}
{\cal D}_\mu = \partial_\mu + i g_{\omega K} V_\mu + 
i g_{\rho K} \vec{\tau}_K \vec{R}_\mu 
\label{eq:min_coupl}
\end{equation}
and the effective mass of the kaon is defined as a linear shift of the mass
term by the scalar field
\begin{equation}
m^*_K = m_K - g_{\sigma K} \sigma 
\quad .
\label{eq:scoupl}
\end{equation}
The form of the interaction as mediated by a scalar ($\sigma$) and vector
($V_\mu$, $\vec{R}_\mu$) meson fields is in close analogy to the 
relativistic mean-field model which will be used for the baryon-baryon
interactions. We will focus now on the K$^-$.
The combined equations of motion for the meson fields including nucleons
\begin{eqnarray}
m_\sigma^2 \sigma + b\,m_N (g_{\sigma N} \sigma)^2 
+ c\,(g_{\sigma N} \sigma)^3 &=&
g_{\sigma N} \rho_s + g_{\sigma K} \rho_K \cr
m_\omega^2 V_0 &=&
g_{\omega N} (\rho_p + \rho_n) - g_{\omega K} \rho_K \cr
m_\rho^2 R_{0,0} &=&
g_{\rho N} \left(\rho_p - \rho_n\right) - g_{\rho K} \rho_K
\label{eq:eomK}
\end{eqnarray}
has a certain simple structure. The scalar densities for nucleons and K$^-$
act as a source term for the scalar field. The vector densities are conserved
and build source terms for the corresponding vector fields. The different
signs for the source terms of the isospin dependent $R_0$ field reflect the
isospin of the hadron. Note, that the kaon scalar and vector density are equal
in our approach as the kaon has spin zero, contrary to the nucleon.  
The dispersion relation for the K$^-$ at zero momentum is given by
\begin{equation}
\omega_K = m_K - g_{\sigma K}\sigma -
g_{\omega K} V_0 - g_{\rho K} R_{0,0}
\quad .
\label{eq:disp}
\end{equation}
The solution of the equations of motion provides then an EoS of the form
\begin{eqnarray}
\epsilon &=& \epsilon_N  + \epsilon_K  + \epsilon_{e,\mu} \\
p &=& p_N + p_{e,\mu}
\quad .
\end{eqnarray}
Note, that the direct contribution of the kaons to the pressure vanishes, as
it involves a Bose condensate. The energy contribution of the K$^-$ reads
\begin{equation}
\epsilon_K = m^*_K \rho_K
\quad .
\end{equation}
Now there are two parameters for the K$^-$, which have to be fixed. 
For the vector coupling constants, we use simple quark counting rules and set 
\begin{equation}
g_{\omega K} = \frac{1}{3} g_{\omega N} \quad \mbox{ and } \quad
g_{\rho K} = g_{\rho N} 
\quad.
\label{eq:iso_coupl}
\end{equation}
The scalar coupling constant is fixed to the optical potential of the K$^-$ at 
$\rho_0$:
\begin{equation}
U_K (\rho_0) =  -g_{\sigma K} \sigma( \rho_0) - g_{\omega K} V_0 (\rho_0)
\end{equation}
and will be varied between $-140$ and $-80$ MeV according to the results of
the coupled channel calculations of the previous section.

\subsection{Phase transition to kaon condensation}

The phase transition to kaon condensed matter can be in principle of any
order. If the phase transition is of first order, then Gibbs' general
condition for two phases have to be applied. As there are two conserved
charges for cold neutron star matter, baryon number and charge, the Gibbs conditions 
read
\begin{equation}
p^{\rm I} = p^{\rm II} \,, \quad \mu_B^{\rm I} = \mu_B^{\rm II}\,, \quad
\mu_e^{\rm I} = \mu_e^{\rm II}
\label{eq:gibbscond}
\end{equation} 
in kinetic and chemical equilibrium. Note, that the standard Maxwell
construction can not be used, as it ensures to conserve only {\em one}
chemical potential. 

\begin{figure}
\begin{center}
\includegraphics[width=0.8\textwidth]{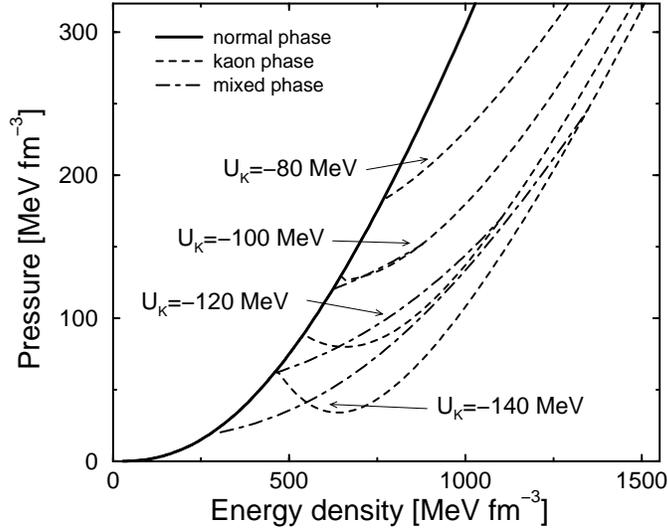}
\end{center}
\caption[]{EoS of kaon condensed matter. The physical solution of the mixed
  phase is given by the Gibbs construction (dash-dotted lines) as taken from \cite{GS99}}
\label{fig:eos_all_bw}
\end{figure}

For a sufficiently large attraction for the K$^-$, $U_K(\rho_0)>-80$ MeV, we
find that the phase 
transition is indeed of first order. Figure~\ref{fig:eos_all_bw} shows the EoS of 
neutron star matter with a kaon condensate. The pure charge neutral nucleon
phase is shown by the solid line. The pure charge neutral kaon condensed phase 
by the dotted lines. A mechanical instability of the latter phase is apparent for some
density range as the slope gets negative, $dp/d\epsilon<0$. The Gibbs
construction for the mixed phase, plotted by the dash-dotted lines, is mechanical
stable and is a continuously rising function of the density. 
The mixed phase starts at a lower density compared to the
onset of the pure charge neutral kaon condensed phase and can extend to rather 
large densities. The phase transition turns out to be of second order for
small values of the optical potential, i.e.\ $U_K(\rho_0)\geq -80$ MeV.
For both orders of the phase transition, the EoS is considerably softened due
to kaon condensation. 

If  a mixed phase is formed, there is a new degree of freedom to
maximize the pressure: the redistribution of charge between the two 
phases. 
There are three possible solutions for the charge density: i) the pure nucleon 
phase with $q_K=0$, ii) the pure kaon condensed phase with
$q_N=0$, and iii) the mixed phase with 
\begin{equation}
q_{{\rm total}}= (1-\chi)q_N(\mu_B, \mu_e) +\chi q_K(\mu_B, \mu_e) = 0\,,
\end{equation}
where $\chi$ is the volume fraction of the two phases.
The total global charge is still zero, while the two phases
of the mixed phase can have very large local electric (opposite) charge densities.

\begin{figure}
\begin{center}
\includegraphics[width=0.8\textwidth]{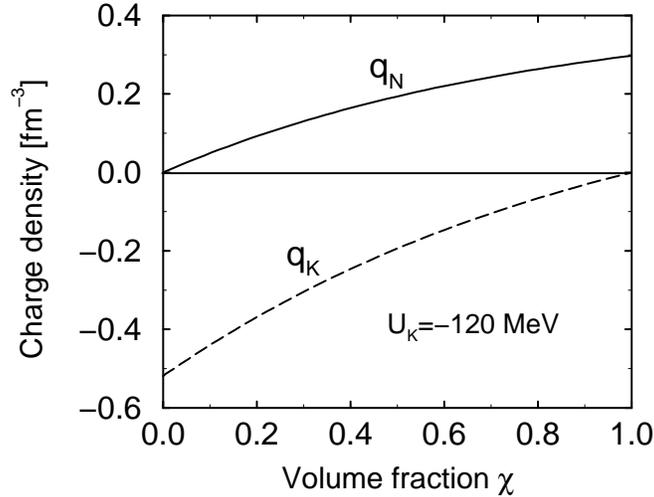}
\end{center}
\vspace*{-0.8cm}
\caption[]{Charge densities of the two phases in the mixed phase (taken from \cite{GS99})}
\label{fig:charge120}
\end{figure}

The charge density in the mixed phase as a function of the volume fraction is
plotted in Fig.~\ref{fig:charge120}. The nucleon phase starts with zero
charge density. Its charge density is getting positive in the mixed phase, as
it is energetically favoured to have about equal amounts of protons and
neutrons in matter. The positive charge of the nucleon phase is compensate by
the kaon condensed phase. The latter phase starts at large negative charge
density and stops at zero charge density at the end of the mixed phase. For
larger density, the pure neutrally charged kaon condensed phase prevails. 
The charge is distributed between the two phases in the mixed phase, and
geometrical (charged) structures appear. These structures are similar to those 
discussed for the liquid-gas phase transition in the neutron star crust 
\cite{RPW83} and for the phase transition to deconfined matter \cite{Glen92}.

\begin{figure}
\begin{center}
\includegraphics[width=0.6\textwidth]{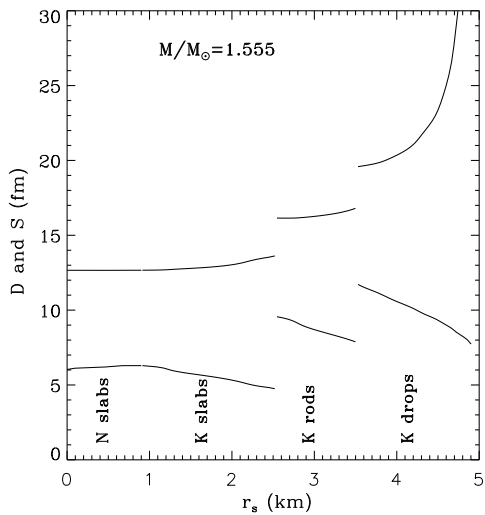}
\end{center}
\caption[]{Sizes of the geometrical structures appearing in the mixed
  phase (from \cite{Christiansen2000})}
\label{fig:structure}
\end{figure}

In the selfconsistent approach used here for nucleons and kaons, the crucial
ingredience for calculating the geometric structures, the surface tension
between the two phases, can be calculated within the model
\cite{Reddy2000,Christiansen2000}. 
The resulting sizes of the geometric structures are summarized in
Fig.~\ref{fig:structure}. First, bubbles of the kaon condensed phase
appear. Then at larger density, here lower radii, the kaon condensed phase forms rods, then
slabs. In the core, the situation reverses and nucleonic slabs form which are immersed
in the kaon condensed phase. If one increases the density even further,
nucleonic rods, then drops will form ending finally in a pure kaon condensed
phase. For the EoS used in Fig.~\ref{fig:structure}, the maximum density
reached inside the neutron star is too low to achieve these phases. The size
of the structures is around 10 fm and the separation 20--30 fm, not far from
the size of heavy nuclei of say 7 fm. Compared to the size of the neutron
star, the mixed phase structures are microscopic and will effect transport and 
cooling phenomena inside the neutron star (see e.g.\ \cite{Reddy2000,Pons2000}).  
 
One might wonder, why there exists a phase boundary for the nucleons, as they
are present in both phases. In the relativistic mean-field model, the mass of
the nucleons is shifted in dense matter due to the interaction with the scalar 
field. In the calculation for the mixed phase, it turns out, that Gibbs
conditions are satisfied for different field configurations, i.e.\ for
different values of the effective nucleon masses in the two phases. The
difference of the effective nucleon masses is about a factor two in the mixed
phase. Hence, the nucleons in the two phases are indeed distinguishable. 
    
\begin{figure}
\vspace*{-0.8cm}
\begin{center}
\includegraphics[width=0.6\textwidth]{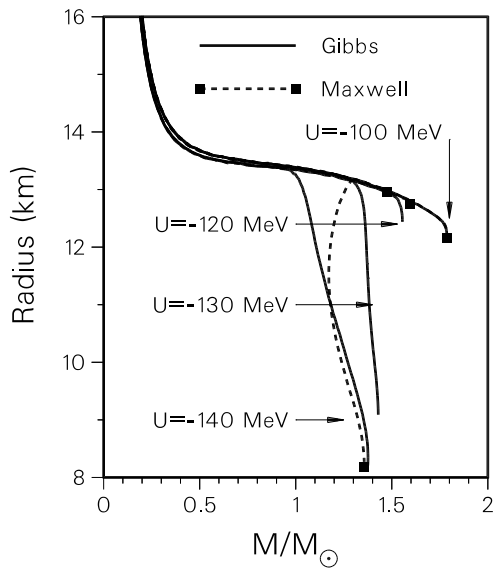}
\end{center}
\vspace*{0.5cm}
\caption[]{The mass-radius relation of a neutron star with kaon condensation for 
  different values of the K$^-$ optical potential at $\rho_0$ (taken from \cite{GS99})}
\label{fig:mrkaon}
\end{figure}

The appearance of the kaon condensed phase has certain impacts also on the
global properties of the neutron star. As kaon condensation softens the EoS,
the maximum mass of neutron stars will be reduced (see e.g.\ \cite{Li97}).
Figure~\ref{fig:mrkaon} shows the mass-radius relation for a neutron star with 
a kaon condensed phase for different strengths of the kaon-nucleon
interactions. For moderate attraction for the K$^-$, the maximum mass of a
neutron stars drops while the minimum radius increases slightly. For larger
attraction, a considerable fraction of the neutron star is in the mixed phase
and the maximum mass as well as the minimum radius decreases. For the largest
attraction studied here, the minimum radius changes drastically from values of 12 km
without K$^-$ condensation to 8 km with K$^-$ condensation. The maximum mass
of the neutron star is lowered from 1.8 M$_\odot$ to 1.4 M$_\odot$. 
Note, that there are no instabilities for the Gibbs construction of the mixed
phase. The Maxwell construction, shown by dashed lines, is mechanically unstable 
for some intermediate ranges of the radius.

\subsection{Effects of hyperonization on kaon condensation}

Hyperons may appear around twice normal nuclear density in beta-stable matter
(see \cite{Glen85} and references therein). In the last few years, this
picture has gained support by various different, modern model
calculations. Hyperons (either the $\Lambda$ or the $\Sigma^-$) are
present in neutron star matter at $2\rho_0$  
within effective nonrelativistic potential models \cite{Balberg97},
the Quark-Meson Coupling Model \cite{Pal99}, extended Relativistic Mean-Field
approaches \cite{Knorren95b,SM96}, Relativistic Hartree-Fock \cite{Huber98},
Brueckner-Hartree-Fock \cite{Baldo00,Vidana00}, 
and chiral effective Lagrangians \cite{Hanauske00}.
Whether the $\Lambda$ or the $\Sigma^-$ appears first, depends sensitively on
the chosen isospin interaction of the $\Sigma$ hyperons. In any case, these
two hyperons appear around (2--4)$\rho_0$ in the model calculations listed
above, which is before kaon condensation sets in.

In most of the modern EoS, the interaction between the baryons is mediated by
meson exchange. The nucleon parameters are fitted to properties of nuclei or nuclear
matter. Some of the hyperon coupling constants, say the ones to the scalar
field, are fixed by the optical potential as extracted from hypernuclear data
\cite{GM91} (see also our discussion in the previous section). The quark model
(SU(6) symmetry) can be  
used to constrain the hyperon coupling constants to the vector fields. The
latter choice is often relaxed. While the $\Lambda$ coupling constants, and to some extend the
ones for the $\Xi$, can be constrained by hypernuclear data, 
the ones for the $\Sigma$ hyperons can not. Studies of $\Sigma^-$ atoms
indicate a repulsive potential for the $\Sigma$ in nuclei
\cite{Mares95}. Another point of uncertainty is related with the interaction
between hyperons themselves. Apart from the $\Lambda\Lambda$ interaction,
nothing is known about the hyperon-hyperon interaction. Only a few calculations
have addressed this issue for neutron star matter so far
\cite{SM96,Balberg97,Vidana00}. 
    
\begin{figure}
\vspace*{-0.5cm}
\begin{center}
\includegraphics[width=0.8\textwidth]{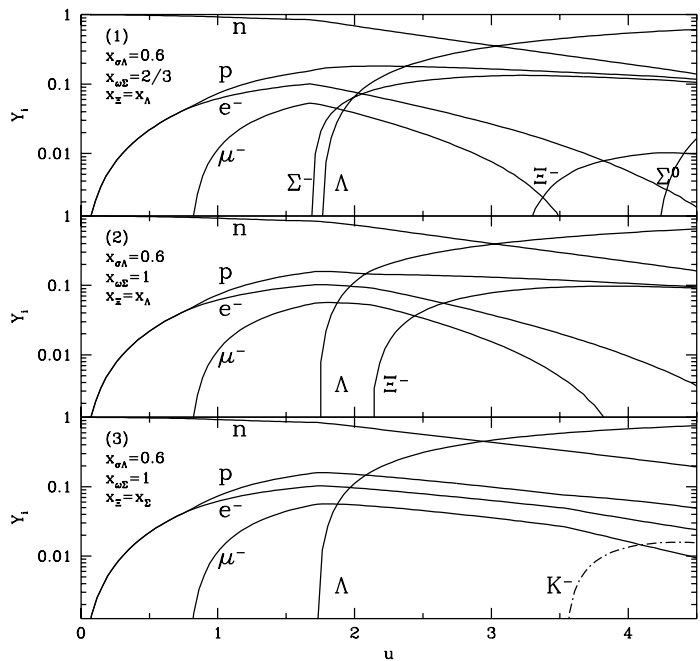}
\end{center}
\vspace*{-0.5cm}
\caption[]{The particle fractions in neutron star matter versus the relative
baryon density $u=\rho/\rho_0$. The three panels correspond to different
choices for the hyperon coupling constants (from \cite{Knorren95b})}
\label{fig:fractions}
\end{figure}

A representative hyperon population as a function of density is plotted in
Fig.~\ref{fig:fractions}. The three panels shown corresponds to three
different choices of the hyperon vector coupling constants. In the upper
panel, the ratio of all hyperon vector coupling constants to the one of the
nucleon is set to about $x_{\omega Y}=2/3$. This is the ratio predicted by the quark model
or SU(6) symmetry for the coupling constant to the $\omega$ meson for the
$\Lambda$ and the $\Sigma$; for the $\Xi$ it would be, of course, only 1/3. 
The $\Sigma^-$ and the $\Lambda$ appear at a little bit less than
$2\rho_0$. The heavier $\Xi$ hyperon is present in matter at
$3.3\rho_0$. If one arbitrarily increases the vector coupling constant of the
$\Sigma$ meson to be equal to the one for nucleons, $x_{\omega\Sigma}=1$, the
$\Sigma$ hyperons do not appear (see middle panel). The critical density for
the onset of the $\Xi^-$ is then shifted to $2.2\rho_0$, so that the $\Xi^-$
takes over to some extent the role of the $\Sigma^-$. If the corresponding
ratio for the $\Xi$ is also increased to $x_{\omega\Xi}=1$, also the $\Xi$
population vanishes (lower panel). The critical density for the $\Lambda$ is
unchanged. Only for the latter case, a kaon condensed phase emerges at
$3.6\rho_0$ in these model calculations. 

Note, that the electron fraction decreases, once hyperons are in the
system \cite{Glen85}. This means, that the electron chemical potential does not only
saturate but is substantially lowered when hyperons are present in matter.  
The obvious reason is, that the negative charge needed to cancel the positive charge
of the protons is carried now by the negatively charged hyperons
instead. Another reason is, that any appearance of a new degree of freedom in
matter lowers the overall Fermi momenta of nucleons and leptons, be it charged 
or not. The latter effect is apparent from the lowest panel of
Fig.~\ref{fig:fractions}, where just the presence of the neutral $\Lambda$
hyperon lowers the electron fraction. The general feature, that hyperons lower 
the electrochemical potential, has been restressed by Glendenning most recently
\cite{Glen2000}.
As the onset of K$^-$ condensation is given by the equality of the effective
energy of the K$^-$ and the electrochemical potential, it is evident, that the 
presence of hyperons at least increase the critical density for kaon
condensation. Glendenning discussed even before the work of Kaplan and Nelson
\cite{KN86} the destructive effect of hyperons for the appearance of kaon condensation in
neutron stars \cite{Glen85}. In more recent works, the in-medium effects for
kaons were incorporated in the model calculations including hyperons, and it
was found that the onset for kaon condensation is shifted to higher density
\cite{Knorren95b} or does not take place at all \cite{SM96} (see below).

The hyperonization effects the mass-radius relation for neutron
stars. As for any new degree of freedom, the EoS is also softened by hyperons. 
The maximum possible mass of a neutron star can be lowered by hyperons by about
(0.4--0.7)$M_\odot$ \cite{GM91,Knorren95b}. 
According to the work of \cite{Knorren95b}, kaon condensation without hyperons
reduce the maximum mass by only (0.1--0.2)$M_\odot$ and  
the combined effect of kaons and hyperons shift the maximum mass down by about
0.5$M_\odot$. So, the main effect in the reduction of the maximum mass stems
from the hyperon degree of freedom.
We note, that the reduction of the maximum mass due to kaons only is rather
model dependent. Other estimates find changes up to $\Delta
M=0.4M_\odot$ due to kaon condensation \cite{Li97,GS99}, but hyperons are
ignored in these latter calculations.
 
\begin{figure}
\vspace*{-0.5cm}
\begin{center}
\includegraphics[width=0.6\textwidth]{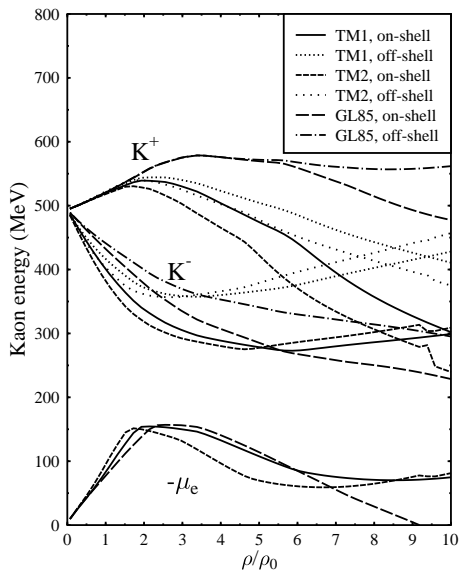}
\end{center}
\vspace*{-1cm}
\caption[]{The effective energy of kaons as well as the electrochemical
  potential versus the baryon density for neutron star matter (taken from \cite{SM96})}
\label{fig:kenrg}
\end{figure}

There can be an additional hindrance for kaon condensation which is related 
to the hyperon-hyperon interaction. The vector meson $\phi$ 
controls the interaction between hyperons at large hyperon densities
\cite{SM96}. The inclusion of the $\phi$ meson models also the kaon-hyperon
interaction. As the $\phi$ meson couples solely to
strange quarks in SU(6) symmetry, it is repulsive for hyperons and the K$^-$
and attractive for the K$^+$. For neutron star matter with hyperons, the
$\phi$ meson exchange can saturate the effective energy of the K$^-$, so that
kaon condensation can not happen at all. This effect is depicted in
Fig.~\ref{fig:kenrg} as a function of density. The effective energy of the
K$^-$ drops at low density while the one for the K$^+$ increases. 
At $2\rho_0$, when hyperons are getting populated, the
effective energy of the K$^-$ saturates as the K$^-$ feels the repulsive
contribution from the hyperons. On the contrary, the K$^+$ energy starts to
drop at this density as the K$^+$-hyperon interaction is attractive. The lower 
curves show the electrochemical potential which has to be crossed by the
effective K$^-$ energy for the K$^-$ to 
be present. As the electrochemical potential as well as the K$^-$ effective
energy saturate at large density due to the
presence of hyperons, that crossing does not happen and kaon condensation is
prevented in this scenario.


\section{Summary and Outlook}
\label{sec:summary}
 
We have discussed the elementary kaon-nucleon interaction as derived from
scattering data and the in-medium changes of the K$^-$ as deduced from kaonic
atoms. Various coupled channel calculations using chiral effective
interactions or the boson exchange model with higher order partial waves and
their results for the K$^-$ properties in dense matter have been reviewed.  
The extracted range of the optical potential has been used to study the phase
transition to kaon condensation in neutron star matter. We outlined the effects of a first
order phase transition due to the presence of structures in the mixed
phase. Consequences for the global properties of neutron stars, i.e.\ a
reduced maximum mass and minimum radius, have been addressed.
Finally, we examined the r\^ole of hyperons on the onset of kaon condensation
in neutron stars. 

It is clear from our review, that the discussion about kaon condensation is
far from being complete at present. The topic is a challenge both for
experimentalists as well as for theorists, demonstrating a growing interplay
between traditional nuclear physics, heavy-ion physics, and
astrophysics. For example, the importance of p-wave
interactions also for cold, dense neutron star matter has been stressed, but
further work is needed in that direction. Especially, the precise value of the 
optical potential of the K$^-$ is a crucial ingredience for the neutron star
matter calculation and needs to be pinned down more precisely, be it by the
experimental study of deep lying levels in kaonic atoms, subthreshold kaon
production in heavy-ion collisions at GSI, or by the mass-radius measurement
of neutron stars. 

The inclusion of hyperons for the equation of state is certainly a necessary
one, but there are only a few works which have been devoted to this issue. 
In particular, the hyperon-hyperon interaction as well as the kaon-hyperon
interaction can be important for the onset of kaon condensation and could be
addressed by e.g.\ a chiral SU(3) symmetric model.  A consistent calculation,
incorporating realistic kaon-baryon as well as a baryon-baryon interactions on
the same basis is still missing. 

Last but not least, we point out, that kaon condensation can have impacts on
other facets of neutron stars, like the evolution of proto-neutron stars and
the deconfinement phase transition to quark matter. Concerning the latter
point, the appearance of both, a kaon condensed phase and a quark matter
phase, will lead to a triple point of strongly interacting matter inside a
neutron star. At this point, the three phases, the normal hadronic, the kaon
condensed and the quark matter phase, are in equilibrium. Beyond that point,
the kaon condensed phase and the quark matter phase are forming a mixed phase.
The presence of this triple point might have interesting implications for
transport phenomena in neutron stars.


\section*{Acknowledgments}

We thank all colleagues that have collaborated with us in obtaining the
results presented here, especially M.B. Christiansen, M. Effenberger,
N.K. Glendenning, S. Hirenzaki, V. Koch, T.T.S. Kuo, I.N. Mishustin, Y. Okumura,
E. Oset, A. Polls, H. Toki, L. Tolos. 
AR acknowledges support by the DGICYT project PB98-1247, the Generalitat de
Catalunya Grant 1998SGR-11 and the EU contract FMRX-CT98-0169.
JSB thanks RIKEN, BNL and the U.S. Department of Energy for providing the facilities
essential for the completion of this work. JW thanks the Gesellschaft f\"ur
Schwerionenforschung (GSI), Darmstadt, and the Bundesministerium f\"ur Bildung 
und Forschung (BMBF) for their support.


\end{document}